\documentclass[aip,preprint]{revtex4-1}

\begin{document}


\title[Measuring thickness in thin NbN films]{Measuring thickness in thin NbN films for superconducting devices}

\author{O. Medeiros}
 \altaffiliation[Also at ]{Department of Interdisciplinary Engineering, Wentworth Institute of Technology }
\author{M. Colangelo}%
\author{I. Charaev}
\author{K. K. Berggren}
\email{berggren@mit.edu}
\affiliation{Massachusetts Institute of Technology, Department of Electrical Engineering and Computer Science,
Cambridge, MA 02139, United States of America}%

\date{\today}

\begin{abstract}
We present the use of a commercially available fixed-angle multi-wavelength ellipsometer for quickly measuring the thickness of NbN thin films for the fabrication and performance improvement of superconducting nanowire single photon detectors.
The process can determine the optical constants of absorbing thin films, removing the need for inaccurate approximations. 
The tool can be used to observe oxidation growth and allows thickness measurements to be integrated into the characterization of various fabrication processes.
\end{abstract}


\maketitle 

\section{Introduction}
Niobium nitride NbN is widely used for the fabrication of superconducting devices such as: superconducting nanowire single photon detectors (SNSPDs) \cite{gol2001picosecond}, 
microwave kinetic inductance detectors (MKIDs) \cite{day2003broadband},
superconducting electronic devices \cite{mccaughan2014superconducting, mccaughan2016using},
nanowire memory devices \cite{zhao2018compact}, 
and hot electron bolometers (HEBs) \cite{khosropanah2007low}. 
The working principle of these devices is intrinsically dependent on the kinetic inductance ($L_k$) which can be related to thickness ($d$) by $L_k \propto d^{-1}$\,\cite{annunziata2010tunable}. 
Measuring the thickness of thin films is therefore critical for the characterization, fabrication, understanding, and improvement of superconducting devices. 

The thickness of a thin film can be measured by a number of mechanical, electrical, and optical methods \cite{heavens1991optical}. 
In each of these categories, the methods rely on some assumption about the material properties. 
Mechanical methods based on accurate mass determination, such as weighing, use the thin film's average density which can be much less than the bulk density and is often difficult to determine \cite{heavens1991optical}. 
Similarly, electrical properties are also highly dependent on the deposition conditions, making any resistance-based thickness determination equally unreliable \cite{heavens1991optical}.
These resistance measurements have also been shown to change over time as a function of surface oxide formation, adding an additional uncertainty source \cite{santavicca2015aging}.
Optically, an infrared transmissiometer has been shown to be a simple and fast method for determining the thickness of NbN thin films \cite{sunter2015infrared}. 
Unfortunately, this technique cannot determine the optical properties (complex index of refraction, optical constants, refractive index and extinction coefficient, $n$ and $k$) of the deposited NbN, leaving a degree of uncertainty in the accuracy of the measurement. 
Other optical methods, such as x-ray reflectivity (XRR), are subject to long alignment and measurement times which makes the measurement difficult to integrate into most fabrication processes.
Alternatively, the thickness can be approximated by the deposition time. 
However, these approximations are generally inaccurate as the deposition rate can vary depending on the deposition parameters and target erosion \cite{kelly2000magnetron}. 
Topographical methods, such as atomic force microscopy or profilometry, can produce high-resolution measurements but require additional fabrication (e.g. etch or lift-off) when determining film thickness. 
The preferred method for measuring a thin film in-process would thus be one that simultaneously determines the refractive index, extinction coefficient and thickness.

Ellipsometry is a well-established method of thin film metrology applicable to plasmonic \cite{oates2011characterization}, semiconductor \cite{lim2002dielectric}, and biosensor \cite{striebel1994characterization} applications.
The thickness and optical constants of transparent thin films, such as SiO$_2$, can generally be determined by ellipsometry with high precision, but due to the absorbing nature of NbN these properties cannot be determined from a single-angle, single-wavelength ellipsometry measurement \cite{archer1962determination}. 
However, it is possible to determine these properties by analyzing multiple samples \cite{mcgahan1993techniques}. 
This method is referred to as the multi-sample analysis method and requires a set of measurements from samples with identical optical properties and varying thickness. 
The measured values will follow a unique curve which is determined by the materials optical constants as the thickness of the film increases. 
This approach is significantly less involved than the use of variable-angle spectroscopic ellipsometry \cite{woollam1999overview} thanks to the fixed angle and limited number of wavelengths necessary for applying this method. 
For that reason, the thickness and optical constants of NbN are best determined by the ellipsometry based multi-sample analysis method.

In this paper, we determined the thickness and optical constants of NbN thin films using a fixed-angle spectral ellipsometer (Film Sense FS-1 Multi-Wavelength Ellipsometer).
Traditional ellipsometer parameters, $\Psi$ and $\Delta$, relating to the change in magnitude and phase of the reflected polarized beam, were measured \cite{archer1962determination}.
They were then used to fit generated $\Psi$, $\Delta$ curves for each wavelength as a function of the film's thickness and optical constants. 
The generated data was produced using OpenFilters, an open-source software package for simulating thin film optical models \cite{larouche2008openfilters} and was fit to the measured data using the Marquardt-Levenberg algorithm \cite{marquardt1963algorithm,Fs1manual2017}. 
The thicknesses determined by the fitting were compared to x-ray reflectivity and sheet resistance measurements and the optical properties were used to track changes in the film over time.

\section{Methods}
The thicknesses of NbN films measured by ellipsometry, x-ray reflectivity, and sheet resistance measurements were obtained to assess ability to measure thin films. 
The result of each technique depends on the technical details of the method. 
In this section we describe how samples were prepared, how each method was conducted, and how the ellipsometer can record changes in NbN and surface oxide thicknesses over time. 


\subsection{NbN Deposition} \label{films}
To produce samples for multi-sample analysis, thin films of NbN were deposited on 220\nm\ of thermal oxide \siotwo\ on Si (100) substrates cut into 1\cm\ by 1\cm\ samples from a 4\inch\ wafer.
The depositions were conducted at room temperature by DC reactive magnetron sputtering using an AJA International Inc Orion series system. 
The deposition conditions were determined by Dane \etal to produce NbN films suitable for SNSPDs at room temperature  \cite{dane2017bias}. 

These conditions were controlled by creating an automated process in the AJA Phase IIJ software and were constant for all samples while the deposition time was varied to produce a range of film thicknesses. 
The deposition times ranged from 20\s\ to 200\s\ in increments of 20\s, and also included 400\s\ and 600\s\ depositions. 
These times were chosen to produce thicknesses between 1\nm\ and 30\nm.

\subsection{Spectroscopic Ellipsometry}
To determine the optical constants and thickness of the NbN thin films, each sample was first measured to produce $\Psi$ and $\Delta$.
To measure a sample, an alignment was performed to be within $\pm$ 0.01\degrees\ and the acquisition time was set to 1\s. 
Longer acquisition times did not show an increase in accuracy. 
The alignment and measurement of a sample was completed in less than 30\s\ using the ellipsometer.
Data from each measurement was then assembled in the ellipsometer's software and evaluated using an optical model. 
The optical model was composed of optical layers that were representative of the measured sample.
Here, the optical model included a \textit{n} and \textit{k} layer for our undefined NbN on top of a thermally grown \siotwo\ layer on a Si substrate. 
In multi-sample analysis it was assumed that each sample has identical layer parameters other than the thickness of a single layer. 
For this reason, a Nb surface oxide layer was not included during the multi-sample analysis and the \siotwo\ layer was fit as a single thickness for all samples. 
The multi-sample analysis simultaneously produced thicknesses for each sample and the \textit{n} and \textit{k} values of our NbN.

Having defined the optical properties of our NbN, a predefined \nbtwoofiv\ layer \cite{rii} was added to the optical model to investigate the oxidation of NbN. 
NbN exhibits poor oxidation resistance and the formation of a surface oxide presents some doubt in our understanding of the film's material properties which can be problematic for fabrication processes like reactive ion etching \cite{toomey2018influence}. 
The product of NbN oxidation has been shown to be \nbtwoofiv\ without the formation of lower valent oxides in earlier studies by Gallagher \etal \cite{gallagher1983oxidation,gallagher1982}.
By adding a \nbtwoofiv\ layer to the optical model, fixed angle ellipsometry was able to track the oxidation NbN. 
In this experiment, a 200\s\ deposition (8\nm) of NbN was transported in reduced atmosphere ($\sim$0.5 atm) box after being removed from the deposition chamber so the exposure to atmospheric conditions before the initial measurement would be minimized. 
The sample was measured over increasing time intervals to track the progress of oxidation while stored in atmospheric conditions. 
The measurement assumed the \siotwo\ thickness was constant for the full duration of the experiment. The oxidation of NbN was observed by fitting the NbN and \nbtwoofiv\ layers in the optical model.

\subsection{Characterization}
The thickness measurements from the ellipsometer were validated by two additional measurement methods.
In the first method, x-ray measurements were performed using a Rigaku Smartlab X-Ray Defractometer conducting parallel-beam x-ray reflectometry which produced x-rays at a wavelength of 0.1541\nm\ from a copper target. 
The XRR measurements took more than 40 minutes per sample. 
In the second method, the resistance of each film was measured after deposition at room temperature using a four point probe. 
The measured resistance ($R$) can be used to calculate the sheet resistance ($R_s$) by $R_s = R a \pi(\ln(2))^{-1}$ \,\cite{van1958lj}, where $a$ is a geometric correction factor.
The thickness can then be related to $R_s$ by $d = \rho R_s^{-1}$ where $\rho$ is the electrical resistivity and $d$ is the thickness.

\section{Results}
In this work, we obtained thickness measurements from each technique and tracked a sample over a period of 100 days. 
These results are necessary for validating the ellipsometer's ability to measure the thickness of NbN thin films. 
The details provided in this section compare the ellipsometer and XRR measurements and present changes in NbN and \nbtwoofiv\ thicknesses over 100 days.

\subsection{Optical Model Validation}
\begin{figure}[t]
    \centering
    \includegraphics[width=8.5cm]{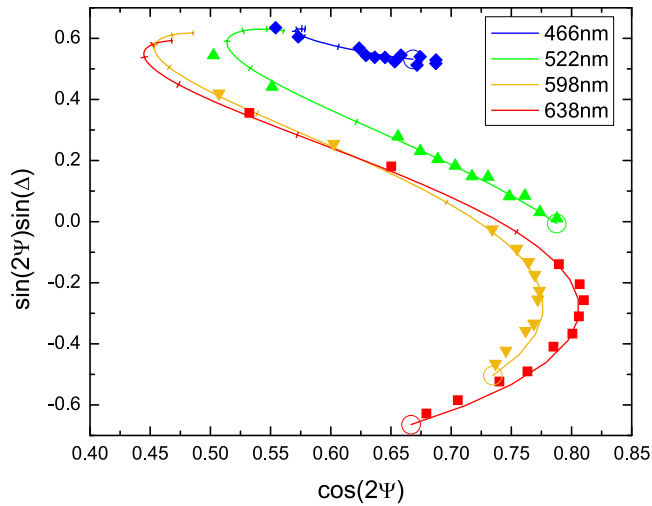}\caption{
    The inferred $\Psi$, $\Delta$ values for each sample are plotted as closed shapes. 
    The legend denotes the corresponding wavelength and its respective color. 
    The curves show simulated data using the optical constants determined by multi-sample analysis. 
    The curves increase from zero thickness (open circles) to 60\nm\ in 10\nm\ increments (tick marks).}
    \label{fig:fig1}    
\end{figure}

The $\Psi$, $\Delta$ curves in Fig. \ref{fig:fig1} shows the locus of increasing thickness for the determined complex index of refraction by fitting to the measured points.
The experimental data from a single sample produced four points, one for each wavelength.
The curves were fit to the experimental data with a mean square error of 0.011. 
The mean square error, or fit difference, calculated by the ellipsometer ranged from 0.055 for the thinnest sample to 0.0156 for the thickest sample.
The values in Table \ref{tab:nk} are the optical constants of NbN determined by multi-sample analysis. 
These values are specific to the NbN produced under the conditions described but are comparable to values previously reported \cite{banerjee2018optical}.  
The model was further validated using the built-in validation software.

\begin{figure}[b]
    \centering
    \includegraphics[width=8.5cm]{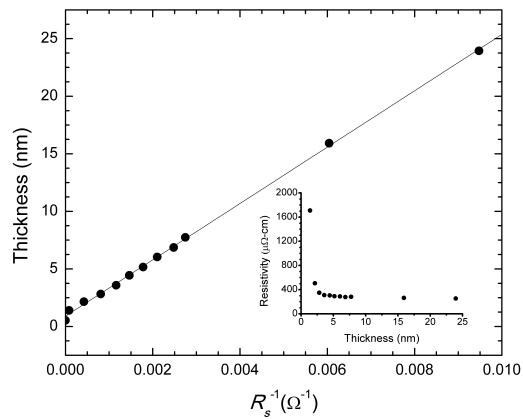}
    \caption{Thickness determined by ellipsometry measurements plotted as a function of sheet conductance. 
    The linear fit shown has a slope of 245\uohmcm. 
    The inset shows the resistivity of the NbN films as a function of thickness.}
    \label{fig:fig2}
\end{figure}

The thickness measurements from the ellipsometer are shown as a function of the sheet conductance in Fig. \ref{fig:fig2}. 
The slope of this linear fit corresponds to a resistivity of 245\uohmcm\ and an adjusted R-square value of 0.9991. 
The fit is expected to have a y-intercept at zero but has an intercept equal to 0.88\nm. 
It can be expected to be zero because zero conductive material should equate to zero thickness.
The resulting intercept implies that resistivity increases with reduced film thickness and this dependency is shown in the Fig. \ref{fig:fig2} inset. 
This increase has been observed by other groups as well and could be attributed to films where the mean free path is less than the bulk material’s mean free path and the charge carrier density is reduced\cite{chockalingam2008superconducting,kamlapure2010measurement}.

\begin{figure}[h]
\centering
    \includegraphics[width=8.5cm]{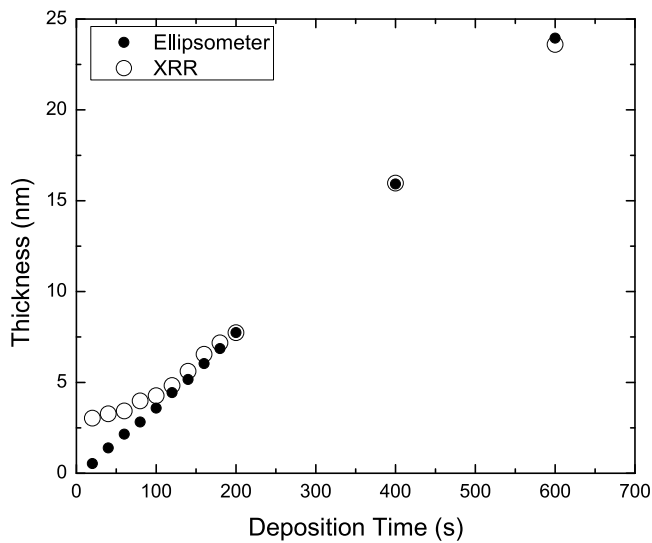}
        \caption{Measured thickness of NbN as a function of deposition time according to XRR and ellipsometer measurements. The XRR measurements are shown as open circles and the ellipsometer measurements are shown as closed circles. As deposition time approaches zero the thickness should approach zero. The XRR measurements for deposition times under 100\s\ deviated from the expected linear trend.}
    \label{fig:fig3}
\end{figure}

A comparison between the measured thickness of each thin film by ellipsometry and XRR as a function of deposition time is shown in Fig. \ref{fig:fig3}. 
In this figure, the XRR measurements for deposition times shorter than 100\s, shown as open circles, deviated from the ellipsometer measurements, closed circles. 
As the deposition time approaches zero, the thickness of the deposited material should also approach zero, which was not observed in the XRR measurements.
However, these measurements on the XRR did not produce interference fringes, likely reducing the accuracy of the fit. 
For film deposition times longer than 100\s\ the ellipsometer thicknesses fell within 0.5\nm\ of the XRR measurements.

\begin{table}[h]
\caption{\label{tab:nk} Optical constants of NbN for each wavelength.}
\begin{ruledtabular}
\begin{tabular}{lll}
$\lambda$ (nm) & $n$ & $k$  \\ 
466 &	2.456 &	2.487 \\ 
522 &	2.554 &	2.536 \\
598 &	2.643 &	3.047 \\
638 &	2.822 &	3.197 \\
\end{tabular}
\end{ruledtabular}

\end{table}

\subsection{Oxide Formation}
\begin{figure*}[h]
    \centering
    \includegraphics[width = 17cm]{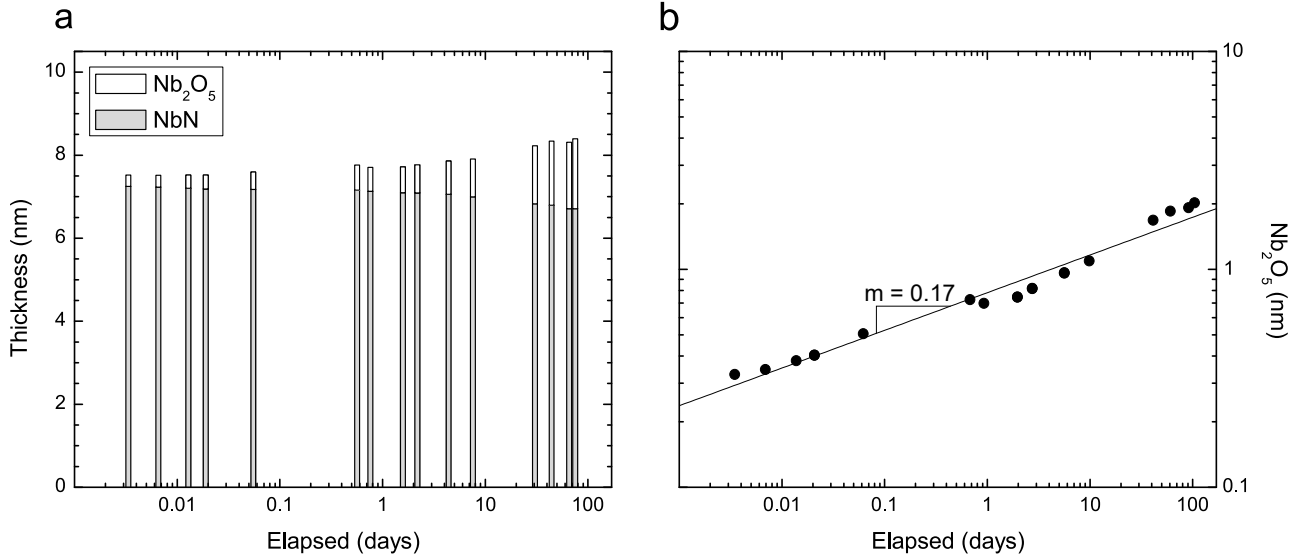}
    \caption{\textbf{a.} Stacked thickness of \nbtwoofiv\ on NbN over a duration of 100 days. The ellipsometer measurements show a decrease in NbN thickness and an increase in \nbtwoofiv\ thickness. The \nbtwoofiv\ is shown in white on top of the NbN shown in gray. 
    \textbf{b.} The same \nbtwoofiv\ thickness measured by the ellipsometer at different intervals after being removed from the deposition chamber is shown with a linear fit.}
    \label{fig:fig4}
\end{figure*}

The time-dependence of thicknesses of NbN and \nbtwoofiv\ according to the ellipsometry measurements are shown in Fig. \ref{fig:fig4}a. 
After 100 days, the measured thickness of the NbN film decreased by 8.7\%.  
The optical model for these measurements used the thickness of \nbtwoofiv\ and NbN as fitting parameters while the \siotwo\ thickness remained constant. 
The \nbtwoofiv\ thicknesses, shown in Fig. \ref{fig:fig4}a and isolated in Fig. \ref{fig:fig4}b, fit the line $d_{ox}=0.78t^{0.17}$ where $d_{ox}$ is the \nbtwoofiv\ thickness and $t$ is the elapsed time in days, with sum of squares error equal to 0.04. 
The oxide formation agrees with the general relationship defined by the Deal-Grove diffusion model \cite{deal1965general}.

\section{Discussion}
The optical method evaluated in this study is a convenient method for simultaneously characterizing the thickness and optical constants of an unknown absorbing film.
Measurements performed using the ellipsometer were, in total, faster than a single measurement performed using the XRR while being more accurate for depositions under 100\s.    
This approach differs from the use of the sheet resistance to thickness relation $\rho=R_sd$ because the ellipsometer is able to provide a qualitative metric for the accuracy of the measurement without relying on an assumption of the material's resistivity. 
By assuming a constant resistivity, it is unknown whether a change in thickness or deposition parameter resulted in a change in a materials sheet resistance. 
Defining the material's optical properties with the ellipsometer can resolve this uncertainty.

Understandably, there are certain scenarios where this method is not applicable.
The first scenario applies to films where there is a strong correlation between the optical constants and thickness. 
This dependency is true for many metallic thin films, including NbN \cite{banerjee2018optical,semenov2009optical}. 
The correlation between thickness and optical constants can depend on thickness as well as the wavelength and crystalline orientation. 
Decorrelating these parameters can be achieved by increasing the number of interference-oscillations. 
This interference enhancement can be accomplished by adding a transparent film below the absorbing layer and simultaneously analyzing multiple samples at various wavelengths or angles of incidence \cite{mcgahan1993techniques,hilfiker2008survey}.
The second scenario occurs when deposition conditions are not constant. 
Changes in the deposition conditions could result in optical properties that differ from those defined by the multi-sample analysis, producing inaccurate results. 
This constraint could be used to determine what effect the deposition conditions have on the structure of the film by observing changes in the measurement's fit difference.  
The third scenario is when the thickness of the film is greater than the film's absorbing limit. 
As thickness of an absorbing film increases the incident beam becomes attenuated and the ellipsometer becomes inaccurate. 
The attenuated beam likely causes the increase in fit difference across our measurements. 
To determine this limit the ellipsometer is able to calculate $\Psi$, $\Delta$ values of an optical model as a function of thickness. 
As thickness increases, the resulting $\Psi$, $\Delta$ coordinate begins to converge to a single value and the measurement is no longer accurate. 
For these NbN films, a conservative upper limit of the measurement is approximately 40\nm.

\section{Conclusion}
We presented a fast, nondestructive and accurate method for determining the film thickness of NbN thin films. 
The results show promise for tracking time or process dependent changes in ultra thin films by determining the optical properties of an unknown absorbing film using a fixed angle multi-wavelength ellipsometer. 
Applications that rely on kinetic inductance or other thickness-dependent electrical properties would benefit from this method. 
By accurately determining the film thickness and optical constants, we construct a better understanding of the material we are creating. 
This understanding can lead to further investigation of the effect that deposition parameters have on optical constants.

\begin{acknowledgments}
The authors would like to acknowledge Blaine Johs for his support with the use of the Film Sense FS-1. They would also like to thank James Daley and Mark Mondol of the MIT Nanostructures Laboratory and Charles Settens of the MIT Center for Material Science and Engineering for their technical support. Support for investigation and characterization of the measurement method itself was sponsored by the Army Research Office
(ARO) and was accomplished under Cooperative Agreement Number
W911NF-16-2- 0192. The views and conclusions contained in this document are those of the authors and should not be interpreted as representing the official policies, either expressed or implied, of the Army Research Office or the U.S. Government. The U.S. Government is authorized to reproduce and distribute reprints for Government purposes notwithstanding any copyright notation herein. Support for the study of oxide growth vs time was provided by Skoltech under Research Agreement No. 1921/R.
\end{acknowledgments}


%
%

%


\bibliography{bib}

\end{document}